\documentclass[aps,prl,twocolumn,showpacs,10pt,superscriptaddress,preprintnumbers,nofootinbib,longbibliography]{revtex4-1}
\usepackage{epsfig,amssymb,amsmath,psfrag,epstopdf,color,siunitx}
\pdfoutput=1
\usepackage{graphicx}
\usepackage[caption=false]{subfig}
 \usepackage[export]{adjustbox}

\usepackage{paralist}
\usepackage{hyperref}
\allowdisplaybreaks

\newcommand{\nn}{\nonumber}

\definecolor{orange}{rgb}{0.9, 0.25, 0}

\definecolor{jdtgreen}{rgb}{0.0, 0.7, 0}

\DeclareRobustCommand{\Fig}[1]{Fig.~\ref{#1}}
\DeclareRobustCommand{\Figs}[2]{Figs.~\ref{#1} and \ref{#2}}

\DeclareRobustCommand{\Eq}[1]{Eq.~(\ref{#1})}

\begin{document}

\preprint{MIT-CTP 5389}

\title{{\color{red}A}nalyzing {\color{red}N}-point {\color{red}E}nergy {\color{red}C}orrelator{\color{red}s} Inside Jets with CMS Open Data}

\author{Patrick~T.~Komiske}
\email{pkomiske@mit.edu}
\affiliation{Center for Theoretical Physics, Massachusetts Institute of Technology, Cambridge, MA 02139, USA\vspace{0.5ex}}
\author{Ian~Moult}
\email{ian.moult@yale.edu}
\affiliation{Department of Physics, Yale University, New Haven, CT 06511, USA \vspace{0.5ex}}
\author{Jesse~Thaler}
\email{jthaler@mit.edu}
\affiliation{Center for Theoretical Physics, Massachusetts Institute of Technology, Cambridge, MA 02139, USA\vspace{0.5ex}}
\author{Hua~Xing~Zhu}
\email{zhuhx@zju.edu.cn}
\affiliation{Zhejiang Institute of Modern Physics, Department of
  Physics, Zhejiang University, Hangzhou, 310027, China\vspace{0.5ex}}

\begin{abstract}
Jets of hadrons produced at high-energy colliders provide experimental access to the dynamics of asymptotically free quarks and gluons and their confinement into hadrons.
In this paper, we show that the high energies of the Large Hadron Collider (LHC), together with the exceptional resolution of its detectors, allow multipoint correlation functions of energy flow operators to be directly measured within jets for the first time. 
Using Open Data from the CMS experiment, we show that reformulating jet substructure in terms of these correlators provides new ways of probing the dynamics of QCD jets, which enables direct imaging of the confining transition to free hadrons as well as precision measurements of the scaling properties and interactions of quarks and gluons.
This opens a new era in our understanding of jet substructure and illustrates the immense unexploited potential of high-quality LHC data sets for elucidating the dynamics of QCD.
\end{abstract}

\maketitle


\emph{Introduction.}---High-energy jets produced at the Large Hadron Collider (LHC) provide a unique opportunity to study the nearly conformal dynamics of high-energy quarks and gluons in Quantum Chromodynamics (QCD) as well as their confinement into hadrons.
The seminal introduction of robust jet algorithms \cite{Cacciari:2005hq,Cacciari:2008gp,Cacciari:2011ma} has enabled detailed measurements of the structure of energy flow within jets, providing a new window into these phenomena.
This in turn has transformed our ability to search for new physics at the LHC \cite{Butterworth:2008iy,Kaplan:2008ie,Krohn:2009th} and offers the opportunity to transform our understanding of QCD itself \cite{Larkoski:2017jix,Marzani:2019hun}.

The study of energy flow in QCD collisions has a long history~\cite{Bjorken:1969wi,Ellis:1976uc,Georgi:1977sf,Farhi:1977sg,Parisi:1978eg,Donoghue:1979vi,Rakow:1981qn}.
Event shape observables were first introduced as resolution variables acting as infrared-safe proxies for the underlying $S$-matrix elements of quarks and gluons.
These observables were well suited for the LEP era where the primary interest was in the distribution of jets themselves, with each individual jet being relatively low energy and consisting of only a few hadrons. 
By contrast, the LHC provides high-statistics samples of individual jets, with high energies ($p_T>500$ GeV) and high particle multiplicities, and the substructure of jets can be measured with remarkable angular resolution~\cite{Kogler:2018hem,CMS:2020poo,ATLAS:2020gwe}.
This massive leap provides an opportunity to rethink the language used for characterizing energy flow in QCD.

\begin{figure}[t]
\includegraphics[width=0.92\linewidth]{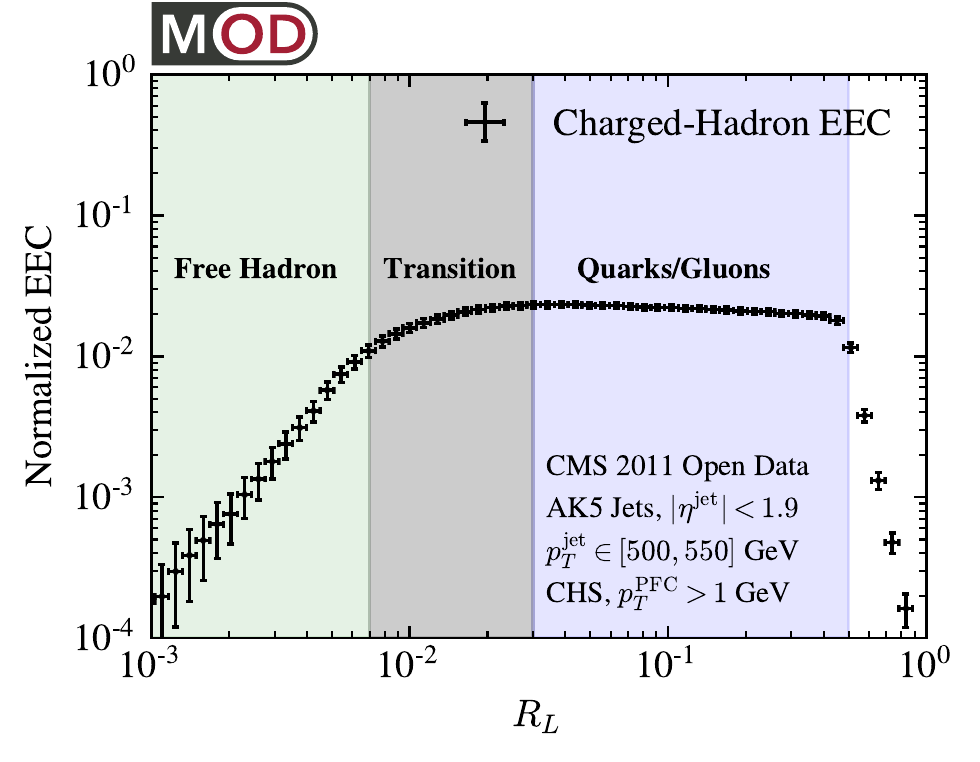}
\caption{The two-point correlator in CMS Open Data, restricted to charged hadrons.
Distinct scaling behaviors associated with asymptotically free quarks/gluons and free hadrons are clearly visible. }
\label{fig:EEC}
\end{figure}

Instead of using shape observables, which take as primary the underlying $S$-matrix elements, it was argued in Ref.~\cite{Hofman:2008ar} that as QCD approaches its conformal limit, one should switch to a characterization of jets in terms of correlation functions.
This enables a beautiful reframing of jet substructure in terms of universal scaling behavior and the operator product expansion (OPE) algebra of light-ray operators. 
Despite the theoretical elegance of the correlator-based approach, measurements of correlators in the perturbative regime require truly high-energy jets, measured with excellent angular resolution, much beyond what was available in the LEP era.
Early studies of these observables in both theory~\cite{Basham:1978bw,Basham:1977iq,Basham:1979gh,Basham:1978zq,Konishi:1979cb} and experiment~\cite{SLD:1994idb,L3:1992btq,OPAL:1991uui,TOPAZ:1989yod,TASSO:1987mcs,JADE:1984taa,Fernandez:1984db,Wood:1987uf,CELLO:1982rca,PLUTO:1985yzc} were thus largely forgotten to history.
With the advent of the LHC, the strong historical preference for jet shapes has left the simplest questions about correlations of energy flow in gauge theories experimentally unanswered.%
\footnote{\Figs{fig:EEC}{fig:EEC_ratio} provide an affirmative answer to Polchinski's question at 47:04 of \cite{MaldacenaKITP}.
We also hope that this introduction provides a historical explanation (although not an excuse!)\ for Maldacena's response: ``People do not do this. I haven't figured out why they don't.''}

To bridge the gap between the real-world environment of QCD at the LHC and theoretical developments in conformal field theory, a program was initiated in Ref.~\cite{Chen:2020vvp} to reformulate jet substructure in terms of correlators.
This program builds on earlier visionary work in the context of conformal field theories~\cite{Hofman:2008ar,Belitsky:2013xxa,Belitsky:2013bja,Belitsky:2013ofa,Belitsky:2014zha,Korchemsky:2015ssa}.
In this paper, we take the next step and use publicly available data released by the CMS experiment to perform the first ever analysis of correlation functions of energy flow operators in high-energy jets.%
\footnote{We use the term ``analysis" instead of ``measurement'' to highlight that we have not corrected the data for detector effects.}
These studies reveal new ways of probing jets at the LHC and transform the beautiful underlying theoretical structures into experimental realities.


\emph{Observables from Correlators.}---Correlation functions are a standard approach to characterizing physical systems, typically building in complexity from simple low-point correlators to more complicated higher-point correlators.
Instead of correlation functions of local operators familiar from condensed matter systems, the objects of interest in collider experiments are correlation functions, $\langle \mathcal{E}(\vec n_1) \mathcal{E}(\vec n_2) \cdots \mathcal{E}(\vec n_k) \rangle$, of the asymptotic energy flow operator~\cite{Sveshnikov:1995vi,Tkachov:1995kk,Korchemsky:1999kt,Bauer:2008dt,Hofman:2008ar,Belitsky:2013xxa,Belitsky:2013bja,Kravchuk:2018htv}:
\begin{align}
\label{energy_flow_operator}
\mathcal{E}(\vec n) = \lim_{r\to \infty} \int\limits_0^\infty dt~ r^2 n^i T_{0i}(t,r \vec n)\,,
\end{align}
where $T_{\mu\nu}$ is the stress-energy tensor.%
\footnote{See Ref.~\cite{Mateu:2012nk} for a variant of the energy flow operator relevant for understanding hadron mass effects.}
These correlation functions (which we refer to generically as EECs) are the fundamental objects of the theory, and are described by an OPE structure~\cite{Hofman:2008ar,Kravchuk:2018htv,Kologlu:2019bco,Kologlu:2019mfz,1822249} that encodes the internal structure of jets.%
\footnote{The positivity of expectation values of \Eq{energy_flow_operator} is an example of an average null energy condition (ANEC) \cite{Tipler:1978zz,Klinkhammer:1991ki,Wald:1991xn,Hofman:2008ar,Faulkner:2016mzt,Hartman:2016lgu}, which pleasingly shares the same initialism as analyzing $N$-point energy correlators.}

Of central physical importance is the scaling behavior of correlators as a function of angular size.
To isolate this feature, Ref.~\cite{Chen:2020vvp} introduced one-dimensional projections of the higher-point correlators obtained by integrating over their shape, keeping only their longest side fixed.
This defines the $N$-point projected correlators:%
\footnote{All observables used in this paper are implemented in publicly available code \cite{EEC_github}.}
\begin{align}
  \label{eq:projection}
\hspace{-0.5cm}  \text{ENC}(R_L) \ = &\ \left(\prod_{k=1}^N \int \!  d\Omega_{\vec{n}_k} \right)
\delta (R_L - \Delta \hat R_L)\\
&\ \cdot \frac{1}{(E_{\rm jet})^N} \, \langle  
{\cal E}(\vec{n}_1) {\cal E}(\vec{n}_2)  \ldots 
{\cal E}( \vec{n}_N)   \rangle \,, \nn
\end{align}
where $d \Omega_{\vec n}$ is the area element on the detector, $\Delta \hat{R}_L$ is an operator selecting the largest angular distance between the $N$ measured directions, and the average is over an ensemble of high energy jets with energy $E_{\rm jet}$.
For hadron collider measurements, we use the standard longitudinally-boost-invariant transverse momentum $p_T$ as the energy coordinate and $\Delta R=\sqrt{\Delta y^2 + \Delta \phi^2}$ in the rapidity-azimuth plane as the angular coordinate.%
\footnote{For those familiar with the discussion of energy correlators in the CFT literature, one should simply associate $\Delta R^2$ with the conformal cross ratio $\zeta$.}
In the perturbative regime, the projected correlators exhibit a single-logarithmic scaling governed by the twist-2 spin $j=N+1$ anomalous dimensions \cite{Chen:2020vvp}.
They therefore capture the scaling properties of a generic $N$-point correlator in a simple one-dimensional observable.

\emph{CMS Open Data.}---Despite being the fundamental objects of the theory, none of these correlators, nor their scalings, have ever been measured at the LHC.%
\footnote{A variant of the EEC using jets instead of individual particles has been measured by ATLAS~\cite{ATLAS:2017qir,ATLAS:2020mee} but due to its use of jets, it is not well suited for studying the small-angle limit.}
Furthermore, to our knowledge, no correlator with $k\geq 3$ has ever been measured at a collider experiment.
Fortunately, the public release~\cite{CERNOpenDataPortal} of research-grade collider datasets by the CMS experiment~\cite{Chatrchyan:2008aa,CMS:OpenAccessPolicy} has enabled a new era of open exploratory studies~\cite{Larkoski:2017bvj,Tripathee:2017ybi,PaktinatMehdiabadi:2019ujl,Cesarotti:2019nax,Komiske:2019fks,Lester:2019bso,Apyan:2019ybx,Komiske:2019jim,Bhaduri:2019zkd,refId0,An:2021yqd,Elgammal:2021rne}, allowing us to analyze these correlators on real data.
We have found the use of Open Data to be essential for extracting a consistent picture for the behavior of higher-point correlators, which are not guaranteed to be accurately described by parton shower generators commonly used to study jet substructure observables.
While official measurements by the experimental collaborations remain the gold standard in the field, we believe that Open Data studies are an essential tool for theorists exploring the frontiers of QCD.

Our analysis is based on a reprocessed dataset of jets culled from the CMS 2011A Open Data~\cite{CMS:JetPrimary2011A} and made public in a simple, reusable ``MIT Open Data'' (MOD) format by Refs.~\cite{Komiske:2019jim,komiske_patrick_2019_3340205}.
These jets, clustered using the anti-$k_t$ algorithm with $R = 0.5$~\cite{Cacciari:2008gp,Cacciari:2011ma}, have transverse momenta $p_T \in [500,550]$ GeV and pseudo-rapidity $|\eta| < 1.9$.
To minimize detector effects, we focus on track-based observables (i.e.~those only using charged particles) for most of this paper, given the excellent track reconstruction performance of CMS~\cite{CMS:2014pgm}, including within jets~\cite{CMS:2012oyn}.
Tracks are easily incorporated into the theoretical description of correlators using track functions~\cite{Chang:2013rca,Chang:2013iba,Elder:2017bkd,Elder:2018mcr,Li:2021zcf}.
We identify charged particles from particle flow candidates (PFCs)~\cite{CMS:2017yfk} provided by CMS, which synthesize tracking and calorimeter information.
We follow the procedure in Ref.~\cite{Komiske:2019jim} of using charged hadron subtraction (CHS)~\cite{CMS:2014ata} to mitigate pileup and restricting to PFCs with $p_T > 1$ GeV to minimize acceptance effects.
More detailed studies incorporating detector unfolding will be presented elsewhere.

\emph{Imaging the Confining Transition to Free Hadrons.}---The simplest jet substructure observable is the two-point correlator, which probes the dynamics of a jet as a function of the angular scale $R_L$.
Here, $R_L$ is associated with a transverse-momentum exchange of $\sim p_T^{\rm jet} R_L$ between two idealized calorimeters at infinity.
Since QCD confines, we expect to see two distinct scaling regimes, corresponding to the nearly conformal dynamics of quarks and gluons at large angular scales and to free hadrons at small angular scales. 

In \Fig{fig:EEC}, we show the two-point correlator extracted from the CMS Open Data, which provides a striking confirmation of this picture.
We now describe each region of this plot working from large to small angular scales.
For $R_L \gtrsim 0.5$, the angular size of the correlator is larger than the $R=0.5$ radius of the jet, leading to a behavior that is an artifact of the jet clustering algorithm. 
Moving to smaller angles, we enter a wide regime of universal scaling behavior associated with the perturbative interactions of quarks and gluons, and more explicitly the light-ray OPE and the twist-2 spin-3 anomalous dimensions.
This pristine scaling behavior occurs for over a decade, until at $R_L \sim \Lambda_{\text{QCD}}/p_T^{\text{jet}}\sim 10^{-2}$, there is a clear break in the scaling behavior corresponding to the confinement of quark and gluon degrees of freedom into hadrons.
Below this, we observe a nearly perfect $R_L d\sigma/d R_L \propto  R_L^2$ scaling, corresponding to uniformly distributed hadrons.
Quite remarkably, even if we had no understanding of QCD, we would be able to infer from this analysis that hadrons propagate freely at long distances.\footnote{Strictly speaking, this only shows that energy is uniformly distributed at small angles. We are aware of two ways this can happen:  either there are no interactions or there are infinitely strong interactions \cite{Strassler:2008bv,Hofman:2008ar,Hatta:2008tx}.}

The ability to directly observe a clear transition between interacting partons and free hadrons relies on the high energies of the LHC, where these phases are cleanly separated.
Unlike in condensed matter systems where confinement can be imaged as a function of time \cite{Kormos:2017aa}, one might have naively thought that observing this transition at the LHC would be impossible using only asymptotic measurements.
Fortunately, the time evolution of the jet formation is faithfully imprinted into the angular scale of the correlator, $\tau\simeq1/(p_T R_L^2)$, allowing us to image the jet.\footnote{A music video showing the evolution of the three-point correlator as a function of approximate formation  time can be found at https://youtu.be/ZIFDcAXl73w. This video images the confinement transition in real data from free hadrons at low energies (small angles) to interacting quarks and gluons with non-trivial correlations at high energies (large angles).}
We believe this opens the door to further studies of the confinement transition using LHC data, complementary to the recent Lund plane measurement from ATLAS \cite{ATLAS:2020bbn}, as well as applications to the understanding of the time structure of jet quenching in heavy-ion collisions~\cite{Apolinario:2017sob,Andrews:2018jcm,Mehtar-Tani:2019rrk,Apolinario:2020uvt}.

\begin{figure}[t]
\includegraphics[width=0.92\linewidth]{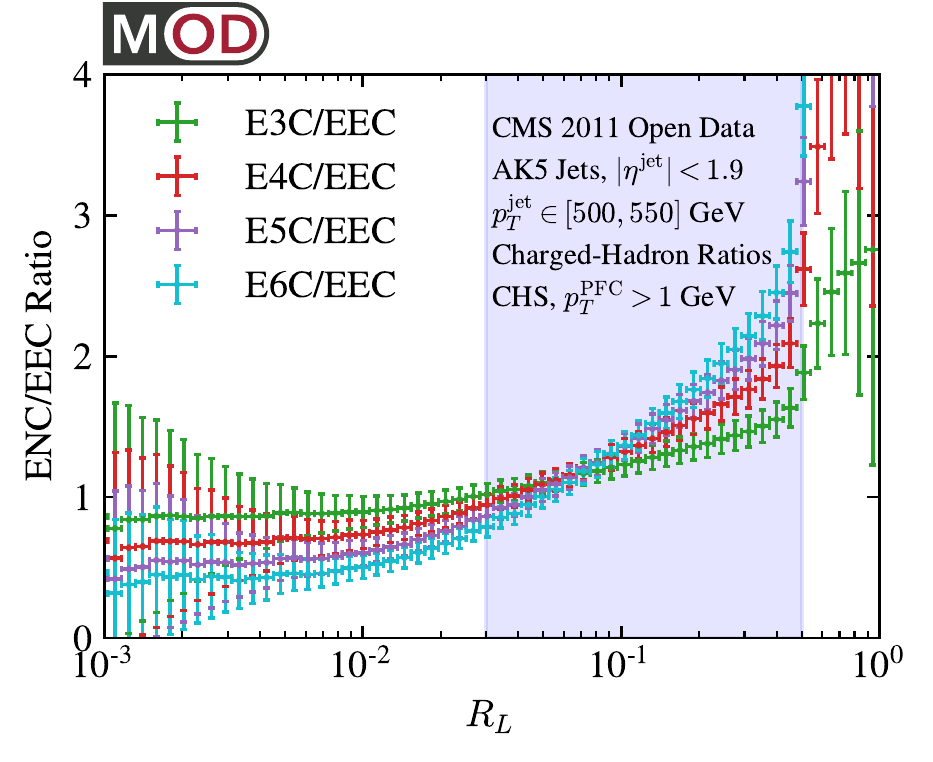}
\caption{
Ratios of the $N$-point projected correlators to the two-point correlator, isolating anomalous scaling in the shaded perturbative regime.}
\label{fig:EEC_ratio}
\end{figure}

\emph{Ratios of Projected Correlators.}---In the wide perturbative window in \Fig{fig:EEC}, the projected $N$-point correlators exhibit a scaling governed by the twist-$2$ spin-$N+1$ anomalous dimensions, providing a precision test of perturbative QCD and a measure of the strong coupling $\alpha_s$ \cite{Chen:2020vvp}.
These correlators have closely related leading non-perturbative corrections for different values of $N$, and thus by taking the ratio to the two-point correlator, we can cancel the leading non-perturbative contribution and isolate a clean perturbative scaling.
Taking the ratio has the added benefit that it removes classical scaling contributions: in the absence of anomalous dimensions, this ratio would be unity.
A non-vanishing scaling in the ratio is therefore a genuine quantum effect associated with the scaling behavior of the light-ray OPE.

In \Fig{fig:EEC_ratio}, we show the ratios of projected correlators up to the six-point correlator.
In the perturbative regime, a clear scaling behavior is observed.
The slope increases as $N$ is increased due to the fact that the twist-$2$ anomalous dimension governing the scaling grows monotonically with spin.
This provides a validation of the predictions of Ref.~\cite{Hofman:2008ar} in public collider data.
Precision measurements of these correlators would be extremely interesting for probing implementations of higher-order DGLAP in parton showers \cite{Hoche:2017hno} and further testing the light-ray OPE.

Additionally, measurements of this scaling behavior provide direct access to $\alpha_s$ and admit a number of advantages over previous proposals to extract $\alpha_s$ from jet shapes.
In particular, this scaling can be measured directly without grooming algorithms \cite{Dasgupta:2013ihk,Larkoski:2014wba}, and can be computed on tracks to significantly reduce experimental uncertainties.
Furthermore, measuring the scaling for a family of projected correlators enables one to disentangle the effects of the parton distribution functions.
We show a comparison of CMS Open Data to leading-logarithmic QCD predictions in the \emph{Supplemental Material}.

\begin{figure}[t]
\includegraphics[width=\linewidth]{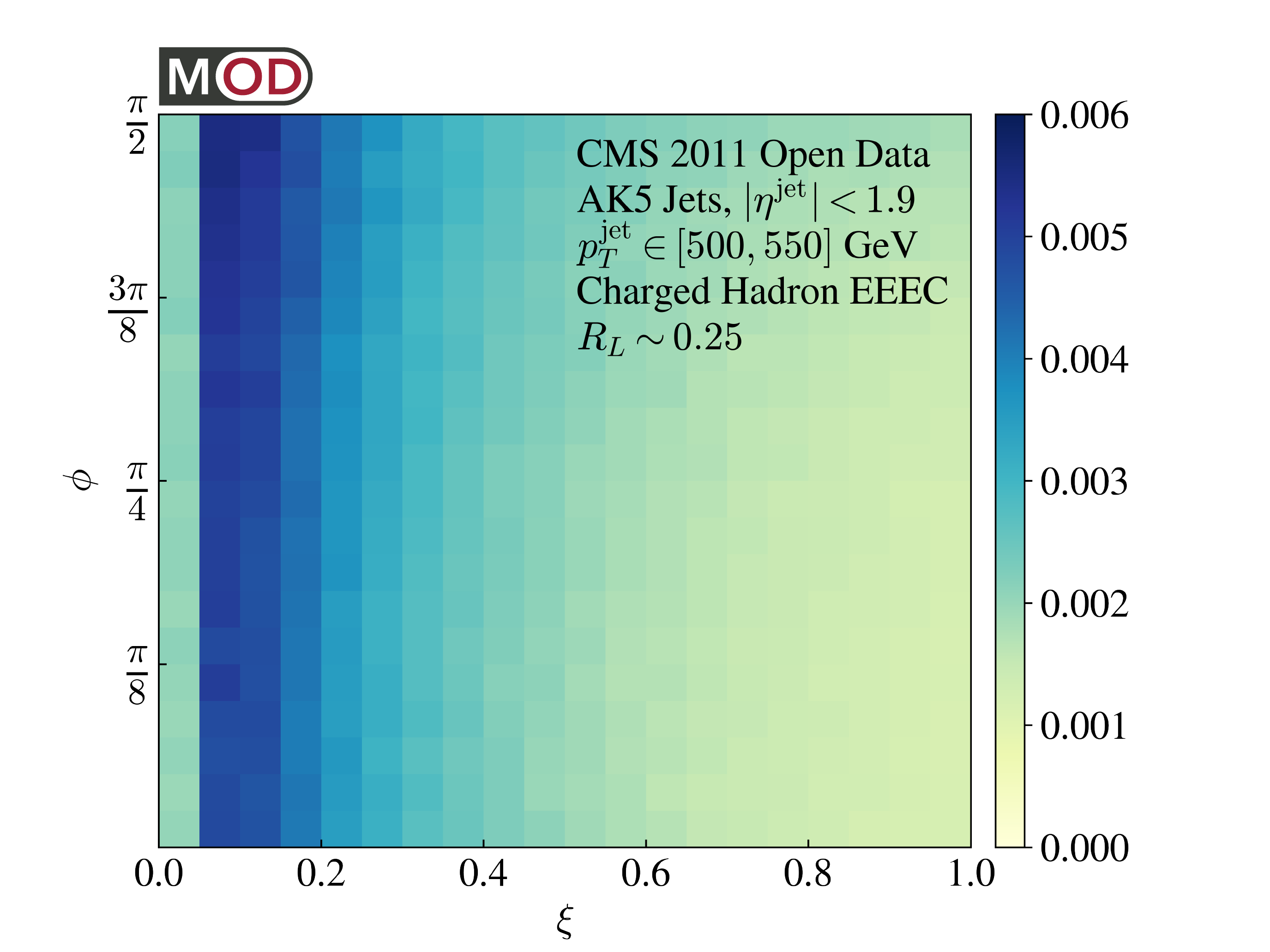}
\caption{The normalized shape dependence of the three-point correlator. Shown here is a slice of the data at $R_L \sim 0.25$ with the coordinates $(\xi,\phi)$ defined in \Eq{eq:transf}.}
\label{fig:EEC_shape}
\end{figure}

\emph{Shapes of Energy Correlators.}---Moving beyond scaling behavior, the shape dependence of higher-point correlators yields insights into the detailed structure of interactions between quarks and gluons.
For example, three-point correlators encode spin correlations~\cite{Chen:2020adz,Chen:2021gdk,Karlberg:2021kwr} arising from the spin-$1$ nature of gluons.
Measurements of higher-point correlators are also useful for testing the incorporation of higher-point splitting functions in parton shower generators.

Here, we focus on the three-point correlator.
For fixed $R_L$, the three-point correlator is a function of two cross-ratios whose analytic form was computed in Ref.~\cite{Chen:2019bpb} to leading order~(LO) in QCD.
For histogrammed analyses, it is convenient to map the domain of definition of the three-point correlation function to a rectangular grid.
Denoting the long, medium, and small sides of the triangle spanned by the operators as $(R_L, R_M, R_S)$, we define the coordinates:
\begin{align}
\xi=\frac{R_S}{R_M} \,, \qquad \phi&=\arcsin \sqrt{1 - \frac{(R_L-R_M)^2}{R_S^2}}
\,.
\label{eq:transf}
\end{align}
This parametrization blows up the OPE region into a line, with $\xi$ and $\phi$ the radial and angular coordinates about the OPE limit, respectively.
More details can be found in the \emph{Supplemental Material}.

In \Fig{fig:EEC_shape}, we show the shape dependence of the three-point correlator in the CMS Open Data, fixing $R_L \sim 0.25$.
It exhibits a rich shape characteristic of the $1\to 3$ interaction in QCD.
This is the first analysis of a three-point correlator in QCD, and more generally, we believe that it is the first experimental analysis of a three-point correlator of light-ray operators in any theory.
The rich LHC data will also enable the measurement of higher-point correlators, as their calculations become available.


\begin{figure}[t]
\includegraphics[width=0.92\linewidth]{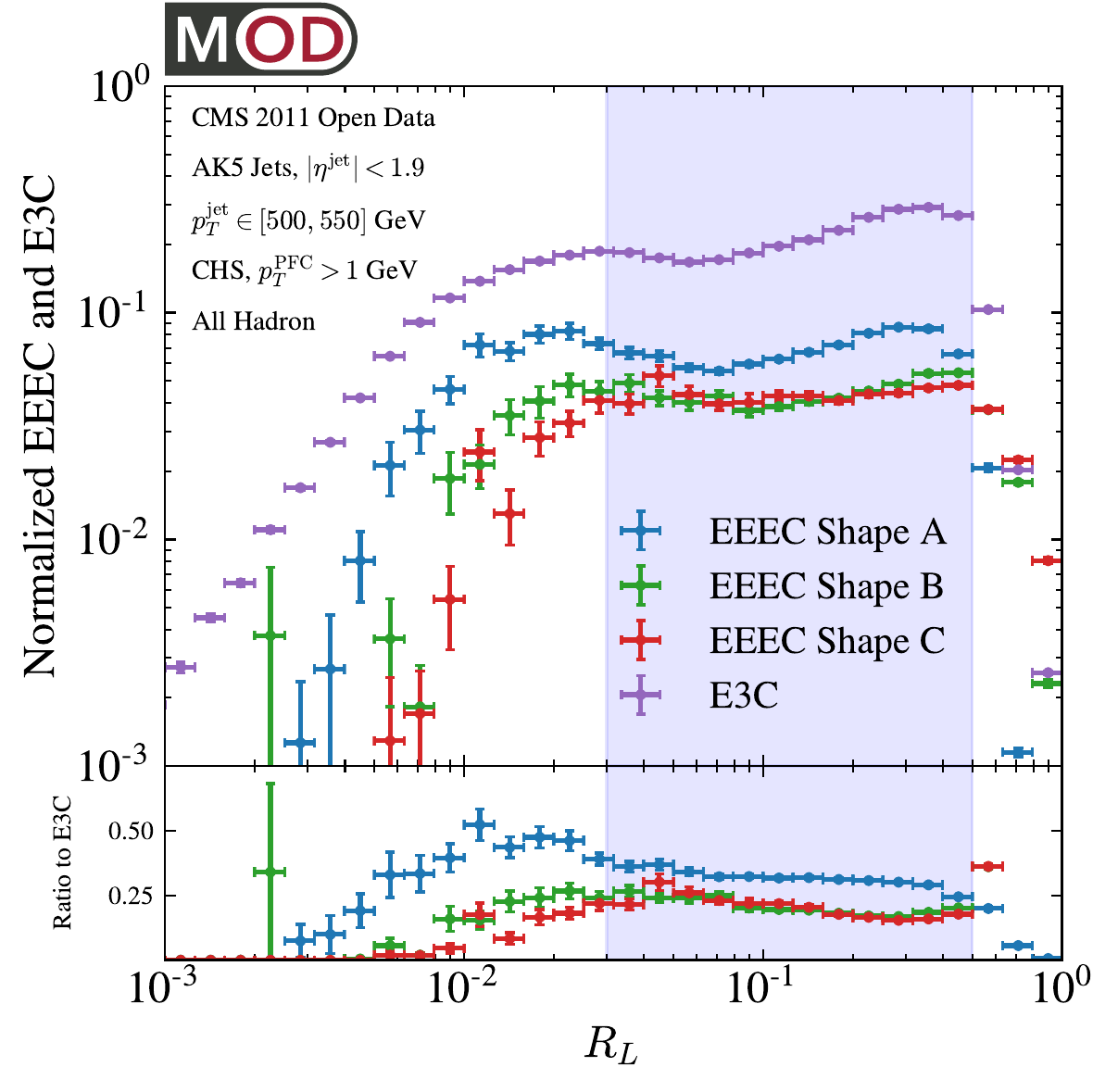}
\caption{Scaling behavior for fixed shapes of the three-point correlator, whose parametrization is given in the \emph{Supplemental Material}. The ratio to the projected three-point correlator is shown in the bottom panel, where flat ratios correspond to the perturbative prediction in the shaded region.  Unlike the previous plots, these results are for all hadrons.}
\label{fig:EEC_shape_scale}
\end{figure}


\emph{Higher-Point Scaling.}---In addition to measuring the shape of the three-point correlator for fixed $R_L$, one can also measure the scaling with $R_L$ for fixed shapes.
One of the remarkable features of the light-ray OPE structure of the energy correlators is that this scaling can be predicted for arbitrary point correlators in conformal field theory~\cite{Hofman:2008ar}.
In the perturbative regime, where the light-ray OPE is applicable in QCD, it predicts that the scaling of an $N$-point correlator of fixed shape is the same as for the projected $N$-point correlator.
This is a much more non-trivial prediction of perturbative QCD, which unlike the projected scaling is not guaranteed to be described by parton shower simulations, making it particularly interesting to study in data.

We focus for concreteness on the scaling of the three-point correlator for fixed shapes. Unfortunately, a LO calculation of the three-point correlator on tracks is not yet available, although it can in principle be obtained using the track function formalism~\cite{Chang:2013rca,Chang:2013iba,Elder:2017bkd,Elder:2018mcr,Li:2021zcf}. We therefore consider only the measurement on all hadrons, though detector effects (which have not been corrected) are larger.
In  \Fig{fig:EEC_shape_scale},  we show the scaling for the three-point correlator measured on all hadrons for three different shapes, denoted by A, B, and C, whose precise parametrization is given in the \emph{Supplemental Material}. The ratio to the projected three-point correlator is shown in the bottom panel. We see consistency with the prediction that the scaling for the shapes is the same as for the projected correlators, though more data and a proper unfolding would be required to make a definitive statement.
Interestingly, as shown in the \emph{Supplemental Material}, this behavior is in tension with the default parton shower in \textsc{Pythia 8.226}~\cite{Sjostrand:2014zea}.
This strongly motivates both more precise measurements of this scaling, and further work to implement the $1\to 3$ splitting functions into parton showers \cite{Li:2016yez,Hoche:2017iem,Gellersen:2021eci}.


\emph{Conclusions.}---In this paper, we argued that taking full advantage of the high energies, multiplicities, and angular resolution of the LHC for studying QCD enables a paradigm shift to thinking about jet substructure in terms of correlation functions of energy flow operators.
Using publicly available CMS Open Data, we showed that the underlying theoretical beauty of the correlator-based approach could be accessible in future experimental analyses, and we illustrated how it provides new perspectives on jets at the LHC.

The focus of this paper has been on the phenomenological applications of correlators to jets at the LHC.
But the rich theoretical structure underlying energy correlators, which has seen remarkable recent progress from numerous directions~\cite{Dixon:2018qgp,Dixon:2019uzg,Chen:2019bpb,Luo:2019nig,Gao:2020vyx,Chicherin:2020azt,Chen:2021gdk,Henn:2019gkr,Kologlu:2019bco,Kologlu:2019mfz,1822249,Korchemsky:2021okt,Korchemsky:2021htm,Chicherin:2020azt,Ebert:2020sfi}, also provides significant motivation for reformulating jet substructure in this language.
This combination of new theoretical techniques and phenomenological applications is truly exciting and opens the door to significant progress in our understanding of QCD using the unique experimental capabilities of the LHC.

\emph{Acknowledgements.}---We are indebted to Cyuan-Han Chang, Hao Chen, Lance Dixon, David Simmons-Duffin, Laura Havener, Petr Kravchuk, Raghav Kunnawalkam-Elayavalli, Yibei Li, Juan Maldacena, Eric Metodiev, Joshua Sandor, Wouter Waalewijn, Meng Xiao, Kai Yan, XiaoYuan Zhang, and Alexander Zhiboedov for many useful discussions, interesting questions, and encouragement.
P.K. and J.T. were supported by the U.S. Department of Energy (DOE) Office of High Energy Physics under contract DE-SC0012567.
I.M. is supported by startup funds from Yale University.
H.X.Z. is supported by National Natural Science Foundation of China under contract No.~11975200.

\bibliography{spinning_gluon}


\newpage

\onecolumngrid

\newpage

\parindent=18pt

\begin{center}
{\large \bf Supplemental Material to \\ {\color{red}A}nalyzing {\color{red}N}-point {\color{red}E}nergy {\color{red}C}orrelator{\color{red}s} Inside Jets with CMS Open Data}
\\
\vspace{3mm}
Patrick T. Komiske, Ian Moult, Jesse Thaler, Hua Xing Zhu\\
\vspace{1mm}
\end{center}

In this \emph{Supplemental Material}, we provide more detailed results on energy correlators from Open Data, simulation, and theory. 

\section{parametrization of Energy Correlators}
\label{sec:parametrization}

\begin{figure}[h!]
\includegraphics[width=0.5\linewidth]{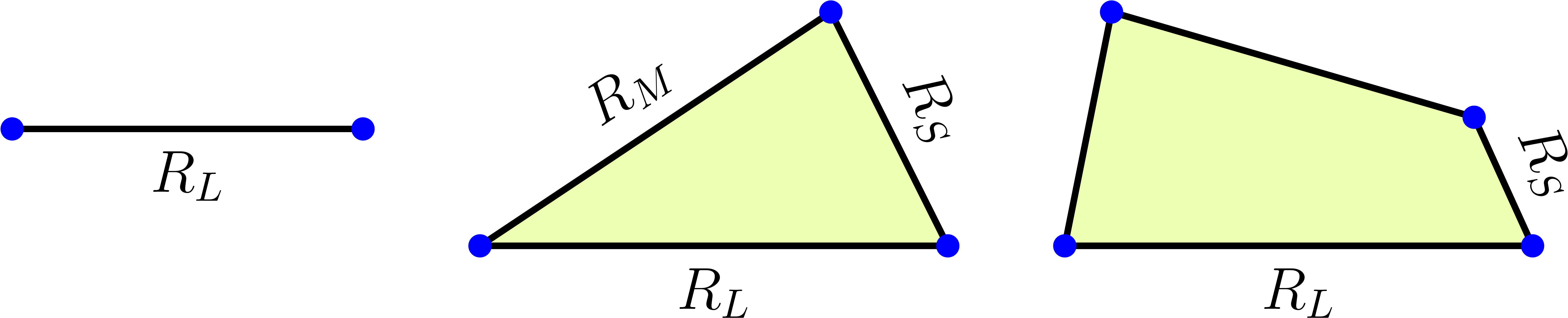}
\caption{Example configurations of two-point, three-point, and four-point energy correlators, labeled by the longest side $R_L$, medium side $R_M$ (in the case of the three-point correlator), and shortest side $R_S$.}
\label{fig:example}
\end{figure}

At hadron colliders, an $N$-point energy correlator is specified by the rapidity and azimuthal angles of $N$ points on an idealized cylindrical calorimeter at infinity.
In the jet substructure (collinear) limit we are considering, they can be well approximated by the configurations of $N$-side polygons (not necessarily convex), with the side lengths specified by the mutual angular distance of the points $\Delta R = \sqrt{\Delta y^2 + \Delta \phi^2}$.
Two-point, three-point, and four-point correlators are shown schematically in \Fig{fig:example}.
A three-point projected energy correlator (E3C) is a three-point energy correlator with $R_S$ and $R_M$ integrated over, while maintaining the hierarchy $R_S \leq R_M \leq R_L$.

\begin{figure}[h!]
\subfloat[]{%
\includegraphics[width=0.255\linewidth,valign=t]{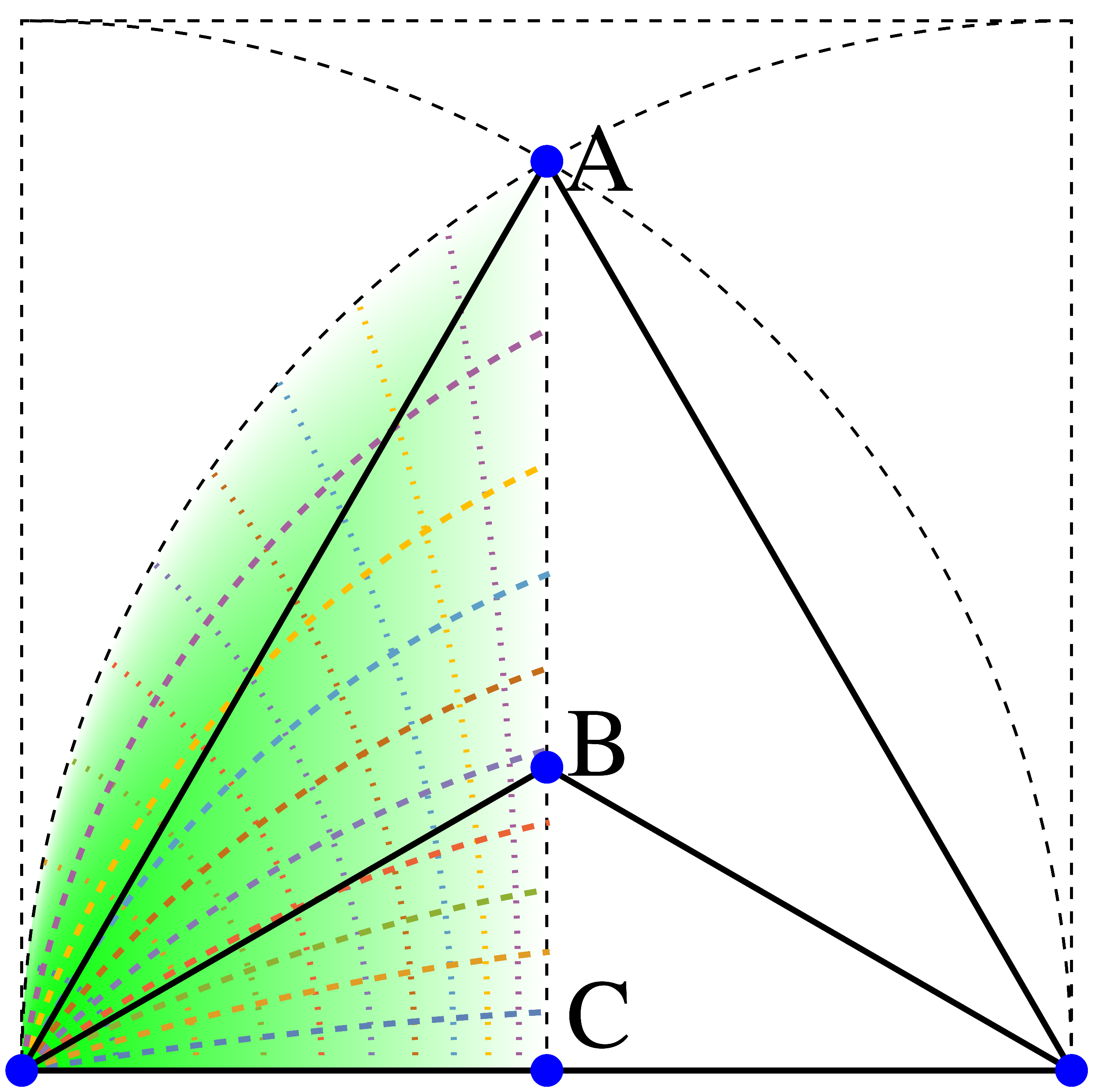}
\vphantom{\includegraphics[width=0.3\textwidth,valign=t]{figs/zzbar_crd.png}}%
\label{fig:conf_zzbar}
}
\qquad\qquad
\subfloat[]{%
\includegraphics[width=0.3\linewidth,valign=t]{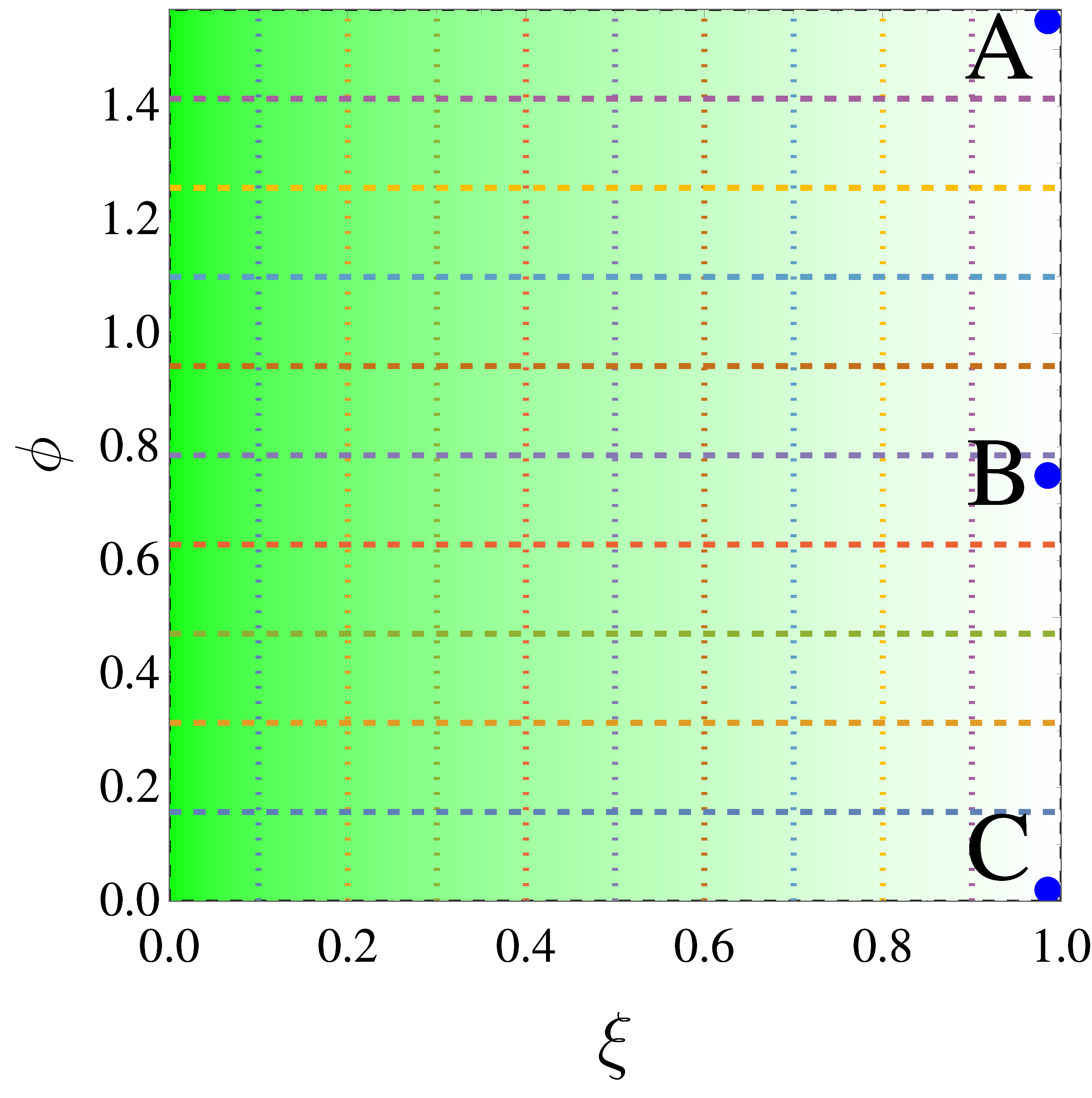}
\label{fig:conf_xiphi}
}
\caption{
(a) Configuration space of triangles with fixed longest side $R_L$.
(b) Mapping the configuration space to a square using the coordinate transformation in \Eq{eq:transf}.  
The OPE singularity is blown up into a line at $\xi = 0$.
The three labeled shapes have $\xi \in [0.975, 1]$, with $\phi$ values of (A) [1.532, 1.571], (B) [0.746, 0.785], and (C) [0, 0.039].
}
\label{fig:configuration}
\end{figure}

In our analysis of the three-point energy correlator, we only distinguish inequivalent configuration up to translation, rotation, and reflection.
We use the configuration space of a triangle to label inequivalent configurations.
This is illustrated by the green region in \Fig{fig:conf_zzbar}.
The squeezed (OPE) limit is located at the bottom left corner.
We also label the three triangles plotted in \Fig{fig:EEC_shape_scale} by A, B, and C in \Fig{fig:conf_zzbar}.
To simplify data binning, we make a coordinate transformation of the configuration space to a square, as in \Eq{eq:transf}.
A schematic illustration of the mapping is shown in \Fig{fig:conf_xiphi}, where the squeezed limit has been blown up into a line at $\xi = 0$.

\section{Comparison with Leading-Logarithmic Predictions}
\label{sec:LL}

It is instructive to compare theory predictions for the $N$-point projected correlator against results obtained from the CMS Open Data.
For simplicity we restrict our theory prediction to leading-logarithmic (LL) accuracy.
In principle, a next-to-leading logarithmic analysis could be carried out using the formalism in Refs.~\cite{Chen:2020vvp,Dixon:2019uzg}, combined with the use of fragmenting jet functions to incorporate the jet algorithm dependence~\cite{Procura:2009vm,Kang:2016ehg,Kang:2016mcy}.
Next-to-next-to-leading logarithmic predictions are also available for $e^+e^-$ collisions~\cite{Dixon:2019uzg}, while for hadronic collisions, an infrared subtraction algorithm for collinear unsafe final state observable is needed, which is currently not available.

In the LL approximation, the $N$-point projected energy correlator is given by the following factorization formula at the factorization scale $\mu$~\cite{Chen:2020vvp}:
\begin{equation}
\text{ENC}(R_L) = \frac{d}{dR_L} \left[ 
(1,1) \exp\left( - \frac{\gamma^{(0)}(N+1)}{\beta_0} \ln \frac{\alpha_s(R_L \mu)}{\alpha_s(\mu)} \right)
\begin{pmatrix}
x_q
\\
x_g
\end{pmatrix}
\right] H_J(\mu) \,,
\label{eq:LL}
\end{equation}
where $\beta_0 = 11 C_A/3 - 2N_f/3$ is the one-loop QCD beta function, $x_q$ ($x_g$) is the fraction of quark (gluon) jets in the sample, and $H_J$ is the production cross section for a jet under the $p_T$ and rapidity selection cut.
Note that at LL, the $N$ dependence only enters through $\gamma^{(0)}(N+1)$.
At leading order, $\gamma^{(0)}(j)$ is the anomalous dimension matrix of twist-$2$ local Wilson operator for quarks and gluons:
\begin{equation}
\gamma^{(0)}(j) = 
\begin{pmatrix}
\gamma_{qq}^{(0)}(j) & 2N_f \gamma_{qg}^{(0)}(j)
\\
\gamma_{gq}^{(0)}(j) & \gamma_{gg}^{(0)}(j)
\end{pmatrix} \,,
\end{equation}
with matrix entries given by
\begin{align}
  \label{eq:QCDAD}
  \gamma_{ qq}^{(0)}(j)&\ = -2 C_F \left[ \frac{3}{2} + \frac{1}{j (j+1)} - 2 (\Psi(j+1) + \gamma_E ) \right] \,,
\nn\\
\gamma_{gq}^{(0)}(j)&\ = -2 C_F \frac{ (2 + j + j^2)}{j (j^2 - 1)} \,,
\nn\\
\gamma_{gg}^{(0)}(j)&\ = -4 C_A \bigg[ \frac{1}{j (j-1)} + \frac{1}{(j+1) (j+2)}
  - (\Psi(j+1) + \gamma_E)  \bigg] 
- \beta_0 \,,
\nn\\
\gamma_{qg}^{(0)}(j) &\ = - \frac{(2 + j + j^2)}{j (j+1) (j+2)} \,,
\end{align}
where $\Psi(z) = \Gamma'(z)/\Gamma(z)$ is the logarithmic derivative of the gamma function.

The expression in \Eq{eq:LL} is a LL prediction at parton level.
At small $R_L$, it scales as $1/R_L$ and is the dominant perturbative contribution.
It is known, however, that EEC-type observables suffer from large hadronization corrections, which scale as $1/R_L^2$~\cite{Basham:1978zq,Korchemsky:1994is,Korchemsky:1997sy,Korchemsky:1999kt}.
When taking the ratio of projected energy correlators, though, a large part of the hadronization corrections are cancelled.
In addition, taking the ratio also largely cancels the hard function $H_J$.
Thus, up to the overall quark/gluon composition, the LL prediction is independent of the parton distribution functions and underlying hard scattering processes that produce the jet ensemble.
This makes the ratio of projected energy correlators an ideal candidate for precision QCD measurements.

\begin{figure}[t]
\subfloat[]{%
\includegraphics[width=0.47\linewidth]{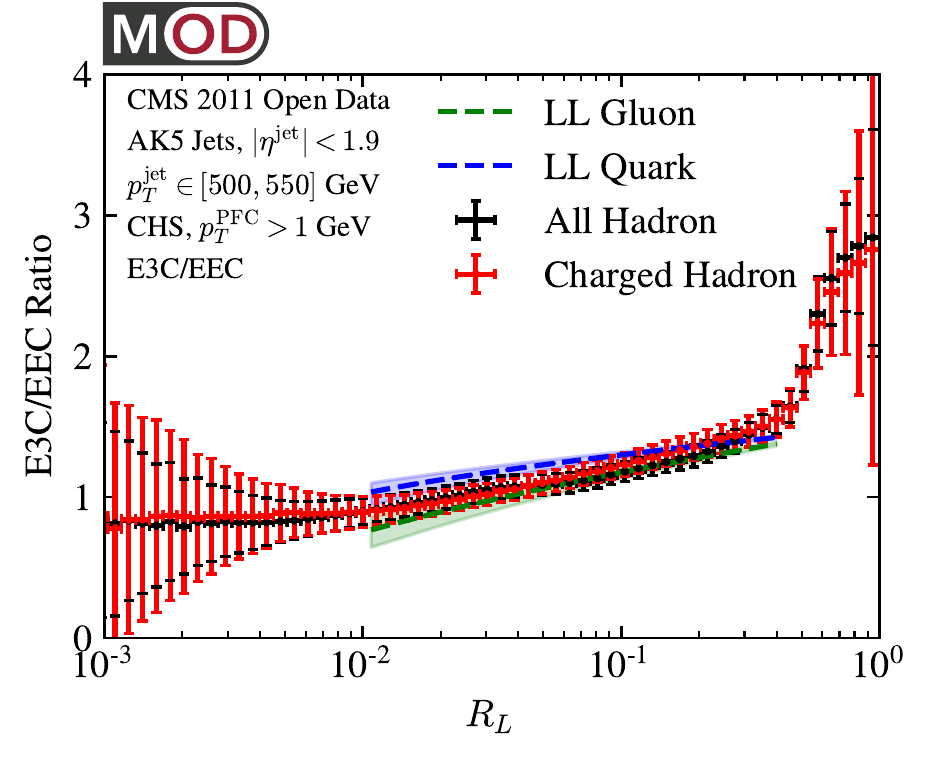}
\label{fig:E3C_ratio}
}
\subfloat[]{%
\includegraphics[width=0.47\linewidth]{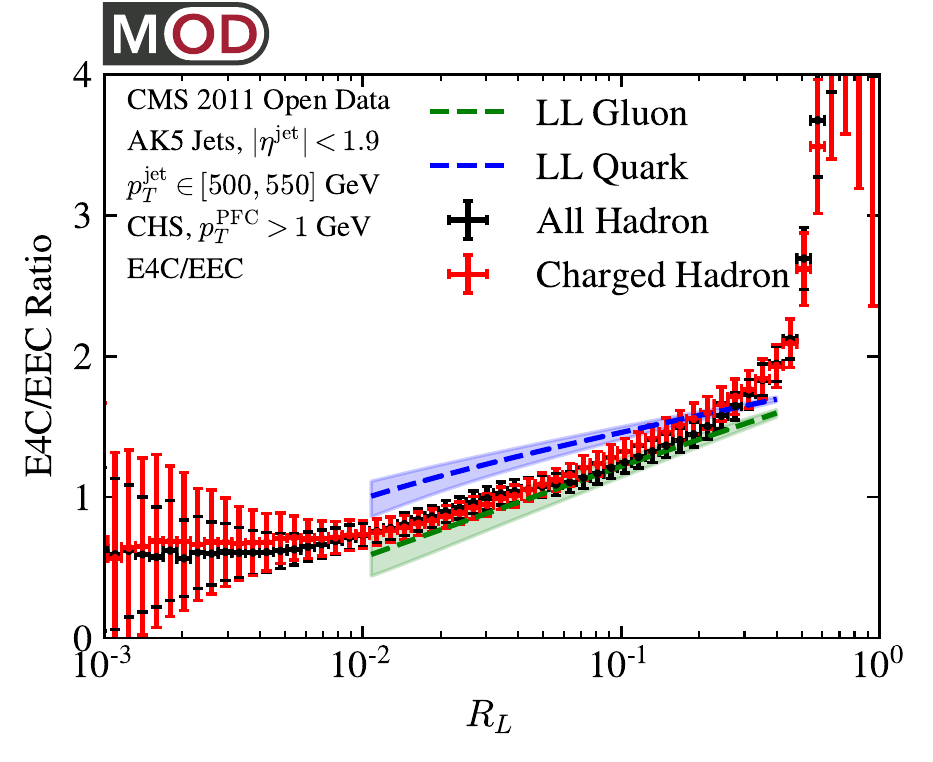}
\label{fig:E4C_ratio}
}
\\
\subfloat[]{%
\includegraphics[width=0.47\linewidth]{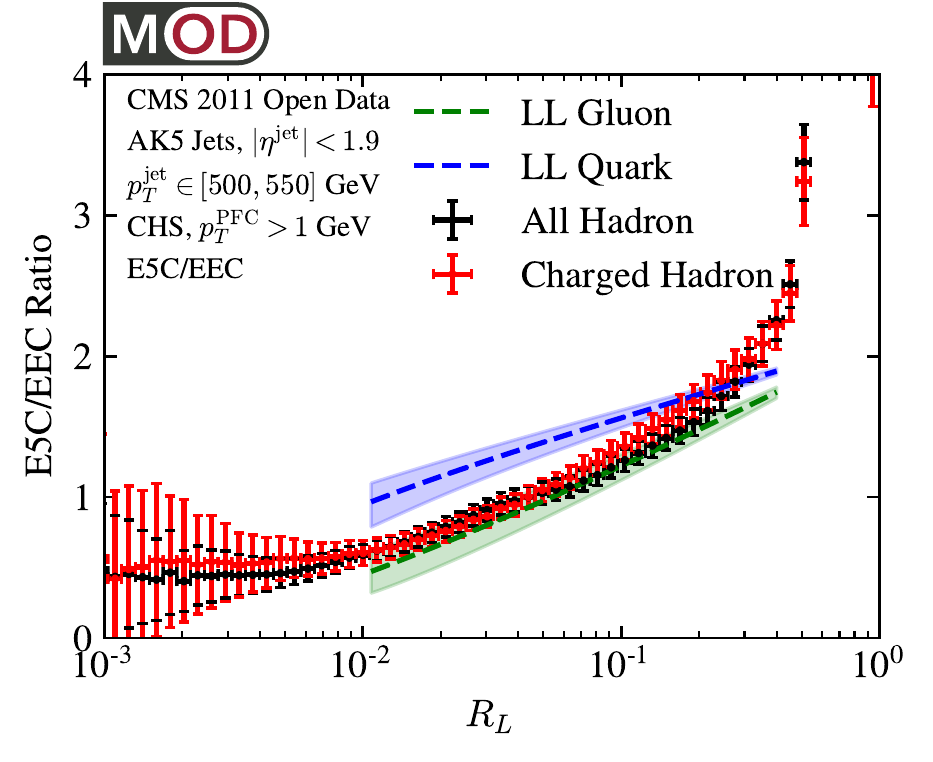}
\label{fig:E5C_ratio}
}
\subfloat[]{%
\includegraphics[width=0.47\linewidth]{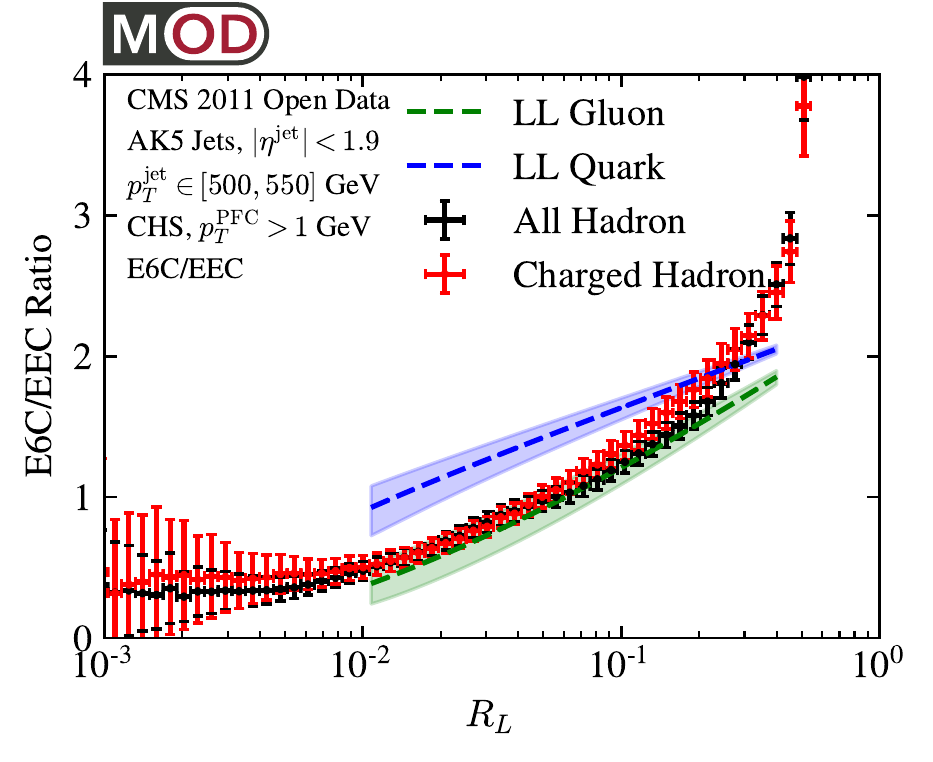}
\label{fig:E6C_ratio}
}
\caption{Ratios of $N$-point projected energy correlators for $N$ ranging from $2$ to $6$.
We show results from the CMS Open Data for both the all-hadron case~(black) and the charged-hadron case~(red).
We stress these results do not involve detector unfolding and the error bars only represents statistical uncertainties.
We also show LL predictions for quark and for gluon jets, with the corresponding scale uncertainty band.}
\label{fig:LL_vs_CMS}
\end{figure}

In \Fig{fig:LL_vs_CMS}, we compare the partonic predictions with CMS Open Data for the ratios of projected energy correlators.
The open data results are shown for all hadrons~(black) and charged hadrons only~(red), and their relative agreement is one piece of evidence for the non-perturbative robustness of these ratios.
The close agreement between the scaling for the ratios of projected correlators as measured on all hadrons and on tracks arises from a combination of three non-trivial features of these observables. First, due to the renormalization group consistency of the hard-collinear factorization formula in \Eq{eq:LL}~\cite{Chen:2020vvp}, the use of tracks does not modify the anomalous dimension of the jet or hard functions. Second, as shown in Ref.~\cite{Chen:2020vvp}, in a pure gluon theory, the track functions are governed by the same anomalous dimensions as the jet function but with a non-trivial mixing structure, leading to an interesting cancellation and resulting in the same LL scaling behavior whether measured on all hadrons or tracks. And finally, corrections to this picture in QCD are suppressed by the difference of the first moments of the track functions for quarks and gluons. Since high energy jets in QCD are dominated by pions, the first moments satisfy the approximate relation $T_g(1)\simeq T_q(1) \simeq 2/3$ and hence $\Delta=T_q(1)-T_g(1)\ll 1$ is highly suppressed \cite{Li:2021zcf,Jaarsma:2022kdd}.
For our LL calculation, we choose $\alpha_s (M_Z) = 0.118$ and use two-loop running of strong coupling. 
We set $\mu =  p_T^{\rm jet}/5$ as the nominal scale, as motivated by the fragmenting jet formalism, and vary around the nominal scale by a factor of $2$ to estimate the theory uncertainty.  
The partonic predictions are shown for a pure-quark sample ($x_q = 1$, $x_g=0$) and a pure gluon sample ($x_q = 0$, $x_g= 1$).
We see that a reasonably good agreement can be achieved if a large gluon jet fraction is chosen.
The fact that good agreement persists out to $N = 6$ is evidence that hadronization corrections are indeed largely cancelled in the ratios. 
In future work, it would be interesting to fit the quark/gluon composition to the data using the technique of Ref.~\cite{Komiske:2018vkc}.

\begin{figure}[t]
\includegraphics[width=0.5\linewidth]{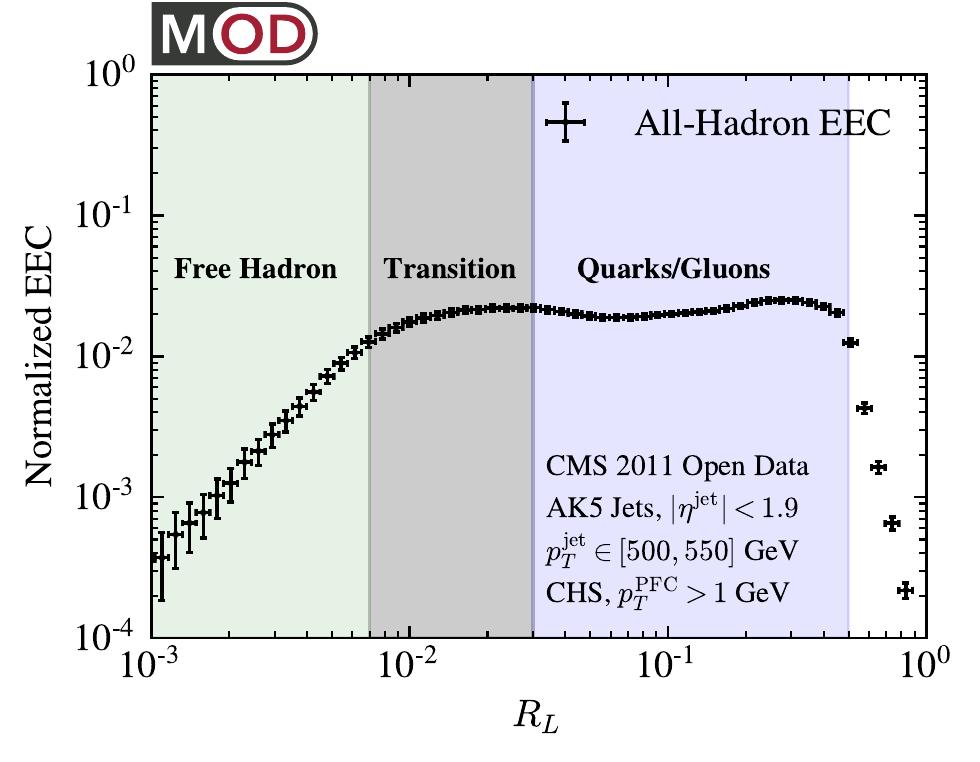}
\caption{EEC from all hadrons in a jet, to be compared to \Fig{fig:EEC}.  No unfolding to particle level is performed, and the uncertainty is statistical only.}
\label{fig:EEC_all}
\end{figure}

For completeness, in \Fig{fig:EEC_all} we show the two-point correlator for the all-hadron case.
Like for the charged-hadron version in \Fig{fig:EEC}, the different phases of QCD are still visible.
That said, in the quark/gluon phase, the scaling law seems to be weakly violated.
We suspect this is due to detector effects, so it will be interesting to see if these features are absent once unfolding is performed.
Perhaps counterintuitively, the $R_L d\sigma/d R_L \propto  R_L^2$ scaling is robust in the all hadron case, despite the worse angular resolution for neutral hadrons.
One has to remember, though, that detector smearing effects also induce decorrelation, so detailed studies are needed to disentangle detector effects from a genuine QCD phase transition.

\section{Comparison with Pythia Parton Shower}
\label{sec:pythia}

\begin{figure}[]
\subfloat[]{%
\includegraphics[width=0.47\linewidth]{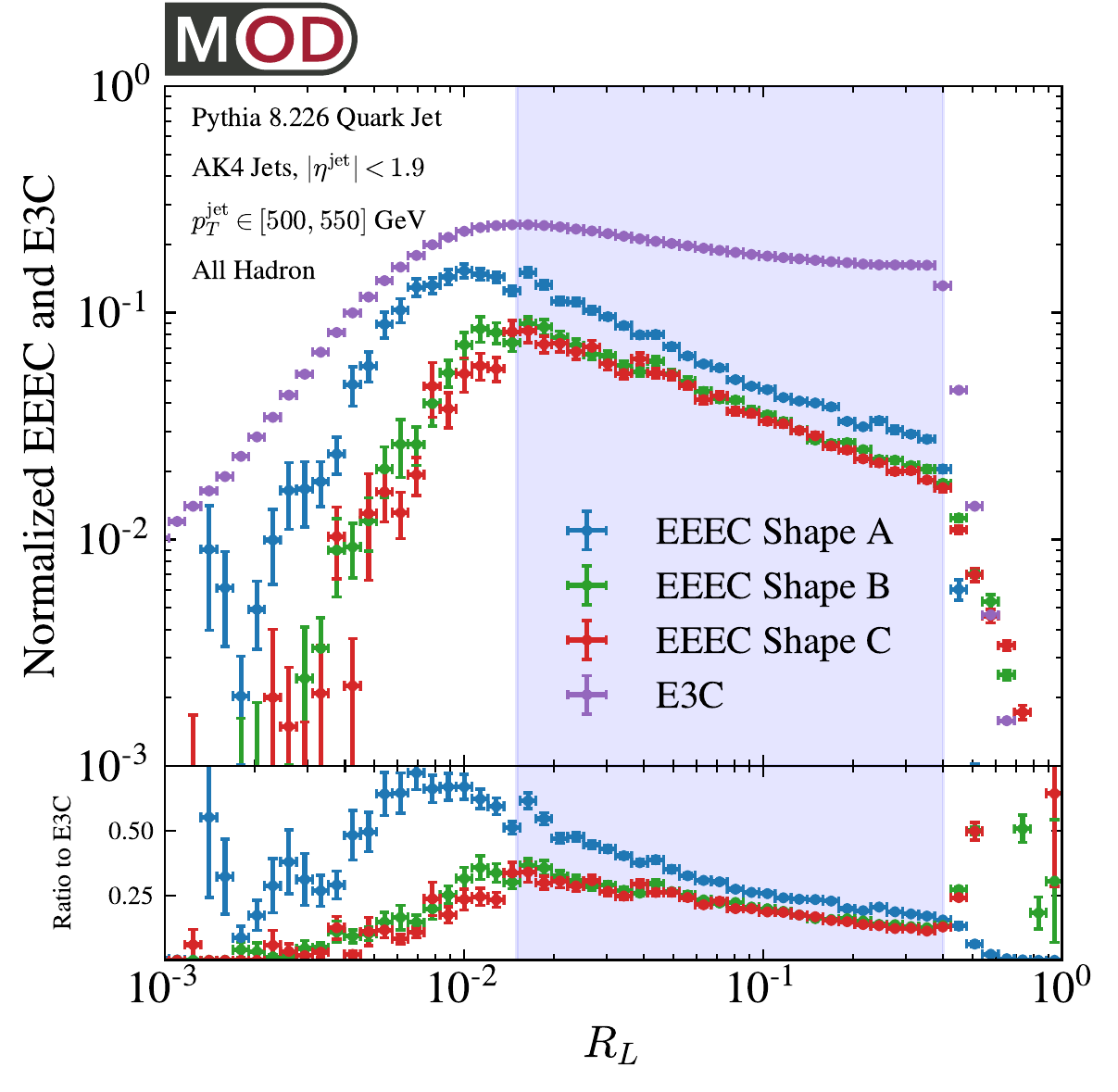}
\label{fig:pythia_quark}
}
\subfloat[]{%
\includegraphics[width=0.47\linewidth]{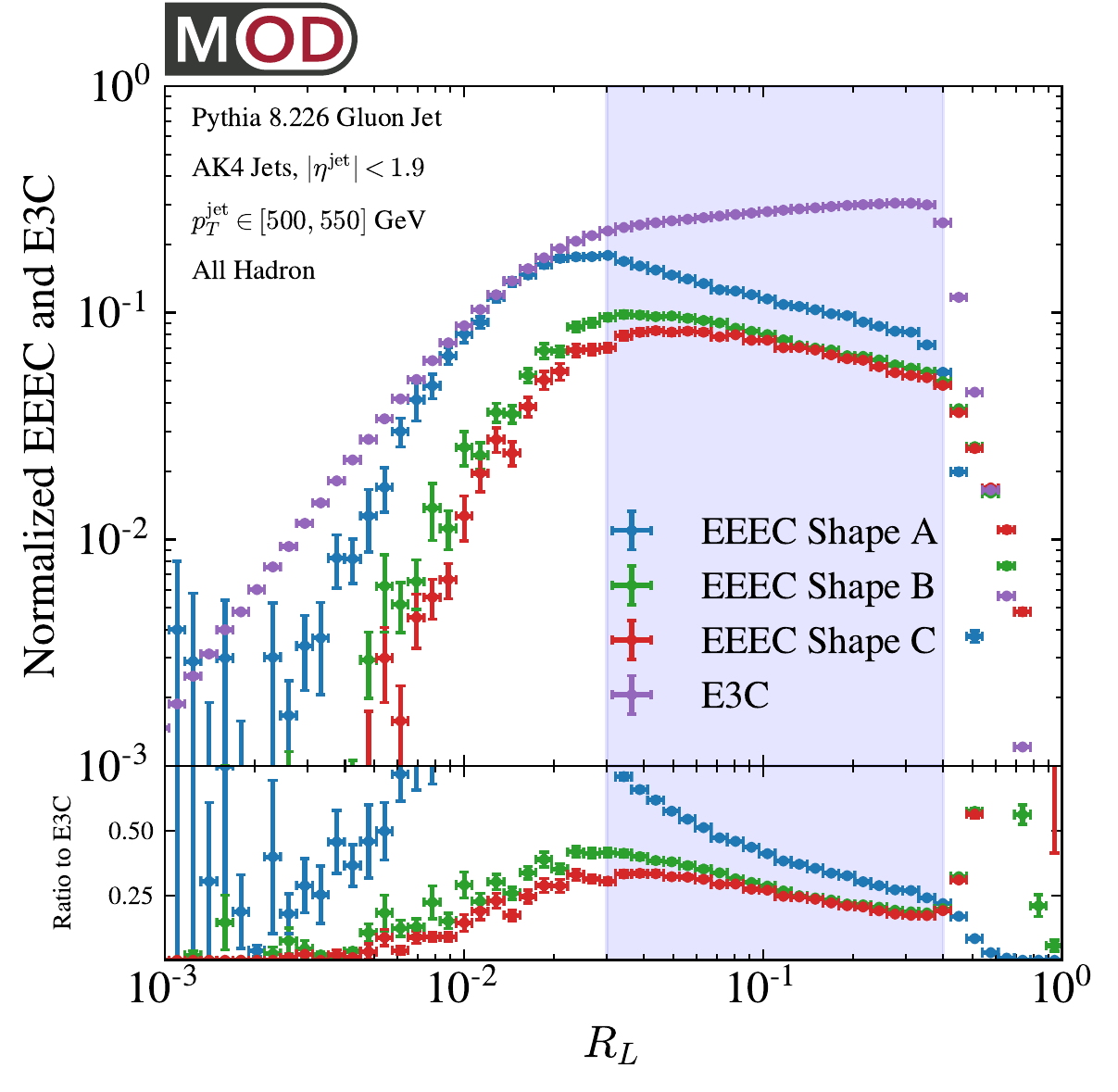}
\label{fig:pythia_gluon}
}
\caption{Three-point scaling behavior in \textsc{Pythia} for fixed shapes for (a) quark jets and (b) gluon jets, to compare to \Fig{fig:EEC_shape_scale}.}
\label{fig:pythia_shape_scale}
\end{figure}

In \Fig{fig:EEC_shape_scale}, we saw that the CMS Open Data results were consistent with theoretical expectations for the $R_L$ scaling of the three-point correlator for different shapes.
For reference, we show results from the default parton shower in \textsc{Pythia 8.226}~\cite{Sjostrand:2014zea}, using quark (uds) and gluon jet datasets generated for the study in Ref.~\cite{Komiske:2018cqr}.
We plot the $R_L$ scaling for three different triangles A~(blue), B~(green), and C~(red), and the normalized projected three-point energy correlator for reference~(purple).
While \textsc{Pythia} does show a scaling behavior for a fixed-shape triangle, the scaling exponent is nevertheless different from the three-point projected energy correlator. This is clearly seen in the bottom panels, where we show the ratio of the scaling for the fixed shapes to the projected correlators.
This seems to contradict the CMS Open Data result in \Fig{fig:EEC_shape_scale}.

In future work, it will be interesting to understand the mismatch between the parton shower and LO predictions.
While \textsc{Pythia} includes LL and partial NLL resummation of logarithms of $R_S$ and the fixed-order prediction does not, the configurations of A, B, and C are chosen such that large logarithms of $R_S$ are less important.
Furthermore, the fixed-order prediction includes matrix-element corrections for $1\to 3$ splittings, whereas the default parton shower in \textsc{Pythia} does not. 
Understanding the origin of difference between Pythia and fixed-order theory, and between Pythia and the CMS Open Data, might shed light on the resummation of $R_L$ scaling and on the matrix element corrections for $1 \to 3$ splitting.

\end{document}